\documentclass[11pt,a4paper,english,superscriptaddress,aps,nofootinbib]{revtex4}
\usepackage{amsmath,amssymb,graphicx}
\makeatletter
\usepackage{babel}
\usepackage[active]{srcltx}
\usepackage{graphicx,color}
\usepackage{changebar}
\usepackage{hyperref}
\hypersetup{
	colorlinks=true,
	urlcolor= blue,
	citecolor=red,
	linkcolor= blue,
	bookmarks=true,
	bookmarksopen=false,
}

\usepackage[dvipsnames]{xcolor}
\usepackage[T1]{fontenc}
\usepackage{ae}
\usepackage[ansinew]{inputenc}
\usepackage{amsmath}
\usepackage{braket}
\usepackage{mathtools}
\usepackage{slashed}
\usepackage{empheq}
\usepackage{tikz}
\usetikzlibrary{decorations.pathmorphing}

\usepackage{graphicx}
\usepackage{amsfonts}
\usepackage{amsmath}
\newcommand{\be}{\begin{equation}}
\newcommand{\ee}{\end{equation}}
\newcommand{\bea}{\begin{eqnarray}}
\newcommand{\eea}{\end{eqnarray}}

\begin{document}
\title{Ground state energy and topological mass in spacetimes with nontrivial topology}
\author{Paulo J. Porf\'irio}
\email[]{pporfirio@fisica.ufpb.br}
\affiliation{Departamento de F\'{\i}sica, Universidade Federal da 
	Para\'{\i}ba,\\
	Caixa Postal 5008, 58051-970, Jo\~ao Pessoa, Para\'{\i}ba, Brazil}

\author{Herondy F. Santana Mota}
\email[]{hmota@fisica.ufpb.br}
\affiliation{Departamento de F\'{\i}sica, Universidade Federal da 
	Para\'{\i}ba,\\
	Caixa Postal 5008, 58051-970, Jo\~ao Pessoa, Para\'{\i}ba, Brazil}

\author{Gabriel Queiroz Garcia}
\email[]{gqgarcia99@gmail.com}
\affiliation{Unidade Acad\^emica de F\'isica, Universidade Federal de Campina Grande, 58429-900,\\
Caixa Postal 10071, Campina Grande, PB, Brazil}

\begin{abstract}
In this paper we study the ground state energy of a massless scalar field and generation of topological mass by considering a quasi-periodically identified Minkowski spacetime and the `half-Einstein Universe', that is, an Einstein Universe where the massless scalar field satisfies the Dirichlet boundary condition. The analysis is performed considering one and two-loop contributions to the effective potential in both cases. We also compare our results with earlier studies.

	\end{abstract}

\maketitle
 \section{INTRODUCTION}
The study of effective potentials plays an important role in quantum field theory as a powerful tool to solve several problems. The effective potential is a function whose minimum, if it exists, describes the vacuum state of the theory. As it is well known, the effective potential can be expanded in power series of $\hbar$, that is, in the form of the loop expansion, so that we can obtain higher-order quantum corrections without losing the information about its classical aspects \cite{Coleman:1973jx}. Such method was proved to be especially useful for the study of vacuum stability and searches for spontaneous symmetry breaking~\cite{Coleman:1973jx, JonaLasinio:1964cw, PhysRevD.35.3187}. Jona-Lasinio in Ref. \cite{JonaLasinio:1964cw}, for instance, applied the effective potential method to understand spontaneous symmetry breaking, and Coleman and Weinberg in Ref.~\cite{Coleman:1973jx} analyzed spontaneous symmetry breaking taking into consideration radiative corrections,  i.e. they studied the dynamical symmetry breaking. In addition, Huang {\it et al.} considered the effective potential for a $\lambda\phi^4$ theory using a $10^4$ lattice~\cite{PhysRevD.35.3187}. Higher-order corrections were investigated in a curved spacetime by Odintsov in~\cite{Odintsov:1993rt} and vacuum stability  issues for a nontrivial spacetime curvature -- by Einhorn {\it et al.} in Ref.~\cite{Einhorn:2007rv}, with focus on non-convex behavior for radiative corrections. The role of effective potentials was also studied in supersymmetric field theories with use of the superfield approach (see for a review ~\cite{Petrov:2001hz} and references therein).

Historically, the first calculation of the effective potential based on using functional methods has been performed by Jackiw \cite{jackiw}. By adopting these methods, Toms studied the Casimir effect (vacuum energy), symmetry breaking and generation of topological mass in Refs.~\cite{toms1, toms2, toms3}. In~\cite{toms1}, for instance, the vacuum energy and generation of topological mass for a scalar field with $\lambda\phi^4$ self-interaction in a periodically identified Minkowski background were discussed. Toms studied in Ref.~\cite{toms2}, by using the $\zeta$-function regularization scheme, loop corrections to the vacuum energy and the topological mass generation for a massless scalar field in spacetimes with non-trivial topologies, including among them, the periodically and anti-periodically identified Minkowski and Einstein spacetimes. In Ref.~\cite{toms3} an interaction between the twisted and untwisted scalar fields was also taken into consideration to study symmetry breaking and topological mass generation.

At the same time, various studies of the Casimir effect, both theoretical and experimental ones, have been performed, see for example \cite{Mostepanenko:1997sw, bordag2009advances, Milton:2001yy}. Even in Minkowski spacetime, when the quantum field is subject to boundary conditions, such as, for example, Dirichlet boundary condition, a nonzero renormalized vacuum energy (the Casimir effect) arises as a consequence of quantum corrections. This phenomenon has been experimentally verified in the case of the electromagnetic field \cite{Sparnaay:1958wg, Lamoreaux:1996wh, PhysRevLett.81.5475, mohideen1998precision, MOSTEPANENKO2000, Bressi:2002fr, PhysRevA.78.020101, PhysRevA.81.052115}. Thus, it was observed that, indeed, when two discharged parallel plates with neglected gravitational interaction are put very close to each other in the vacuum, an attractive force between them takes place as a result of the modification of the quantum vacuum fluctuations. In spacetimes with non-trivial topologies, the nonzero renormalized vacuum energy is straightforwardly affected by the topology of the spacetime \cite{Mostepanenko:1997sw, bordag2009advances}. Recently, in \cite{Bessa:2019aar}, it was shown that the motion of a charged particle over quantum fluctuations of the electromagnetic field requires nontrivial topologies of the spatial section of Minkowski spacetime.

 In this paper we follow the same line of investigation as in Refs. \cite{toms1, toms2, toms3}. We shall consider loop corrections to the vacuum energy of a massless scalar field and generation of topological mass for a quasi-periodically identified Minkowski spacetime and the so called half-Einstein Universe spacetime. Namely, half-Einstein Universe is defined as a scalar field propagating in the Einstein Universe under Dirichlet and Neumann boundary conditions. This spacetime has been considered previously in Ref. \cite{Bayin:1993yq, Ozcan:2001cr, Kennedy_1980}. The quasi-periodic condition has been considered previously in several contexts, for instance in \cite{Feng:2013zza}. Our calculaions, as we shall see,  recover the previous results found in Refs. \cite{toms1, toms2}  in a corresponding limit.

This paper is organized as follows. In section~\ref{secII} we briefly review how to obtain the effective potential by means of the description of path integral method. In the subsequent sub-sections we consider the calculation, at one and two-loop corrections to the effective potential, of the ground state energy and topological mass production in the quasi-periodically identified Minkowski spacetime and the half-Einstein Universe. Finally, in section \ref{secIII} we present our conclusions. Throughout the paper we use natural units $\hbar=c=1$.

\section{Loop corrections and topological mass}
\label{secII}

Let us start this section by the brief review of some aspects of the path integral approach in order to obtain the effective potential. In this sense, a non-minimally {\bf conformally} coupled (to curvature) scalar field theory is described by the following action:
\begin{equation}
S=\int_{\mathcal{M}}\,d^{4}x\,\sqrt{-g}\bigg(\frac{1}{2}(\partial_{\mu}\phi)(\partial^{\mu}\phi)-\frac{1}{12}R\,\phi^2-U(\phi)\bigg),
\label{S}
\end{equation} 
where $\mathcal{M}$ is a Lorentzian spacetime, $g=\text{det}(g_{\mu\nu})$, $R$ is the Ricci scalar, $\phi$ represents a scalar field whilst $U(\phi)$ is the classical potential. {\bf The second term in the former equation is responsible for the conformal invariance of the action when $U(\phi)=0$.} We shall focus our attention on the self-interacting massless scalar field theory $\lambda\phi^4$. In this case, $U(\phi)$ takes the following form
\begin{equation}
U(\phi)=\dfrac{1}{4!}\lambda\phi^4+\frac{C}{4!}\phi^4,
\end{equation} 
where $C$ is the additive counterterm for the coupling constant.

 The next step, as usual,  is to perform a Wick rotation ($t\rightarrow -it$) in the action such that the Lorentzian spacetime ($\mathcal{M}$) is suitably converted into a Euclidean one ($E$).  Taking it into account in the aforementioned action, we turn out getting the Euclidean action, as defined below:
\begin{equation}
S_{E}[\phi]=\int_{E}\,d^{4}x\,\sqrt{g}\bigg(-\frac{1}{2}(\partial_{\mu}\phi)(\partial^{\mu}\phi)-\frac{1}{12}R\phi^2-U(\phi)\bigg).
\end{equation}    

In order to develop a quantum description we shall allow the field $\phi$ to fluctuate around a fixed background field, $\Phi$, with the fluctuations represented by the quantum field, $\varphi$. Thus, one can describe the Euclidean effective action $\Gamma_{E}[\Phi]$ as the generating functional of one-particle-irreducible Green  functions (see Ref. \cite{toms2} for a detailed review): 
\begin{equation}
e^{\frac{\Gamma_{E}[\Phi]}{\hslash}}=\int \mathcal{D}\varphi\,e^{\frac{1}{\hslash}(S_{E}[\Phi+\sqrt{\hslash}\varphi]+J\varphi)},
\label{GE}
\end{equation}
where the integration is performed over all quantum field configurations {\bf and the source is defined by $J=-\dfrac{\delta \Gamma_{E}}{\delta \Phi}$}. Generally, Eq. \eqref{GE} is treated perturbatively by expanding the effective action in power series of $\hslash$ (loop expansion). This also allows us to introduce an effective potential that is written in terms of $\Gamma_{E}[\Phi]$, and that can also be expanded in power series of $\hslash$. Explicitly, we have
\begin{equation}
V_{\text{eff}}(\Phi)=-\frac{1}{\mbox{vol}(E)}\Gamma_{E}(\Phi),
\end{equation}
where vol$(E)$ is the volume of the Euclidean spacetime.  As a consequence, the effective potential can be presented in the form of the loop expansion
 \begin{equation}
V_{\text{eff}}(\Phi)=V_{cl}(\Phi) + V^{(1)}(\Phi) + V^{(2)}(\Phi),
\label{effP0}
\end{equation} 
where $V_{cl}(\Phi)=\frac{1}{12}R\Phi^2+U(\Phi)$  is the tree-level contribution to the effective potential, $V^{(1)}(\Phi)$ and $V^{(2)}(\Phi)$  the one- and two-loop corrections, respectively, and we have taken $\hslash=1$. Note that we have performed a linear expansion about the classical field $\Phi$, i.e., $\phi\rightarrow\Phi+\sqrt{\hslash}\varphi$ \cite{toms2}. Note also that we have only considered quantum perturbations up to two-loop corrections that suffices for our purposes.

The expression for the one-loop contribution to the effective potential is given in terms of the zeta function $\zeta(s)$, that is,
  \begin{equation}
V^{(1)}(\Phi)=-\frac{1}{2\mbox{vol(E)}}\bigg(\zeta^{\prime}(0)+\zeta(0)\log\mu^2\bigg),
\label{effP}
\end{equation} 
 where the prime stands for the derivative of the zeta function with respect to $s$ and the term $\zeta(0)\log\mu^2$ is to be removed by renormalization condition \cite{toms2}. The zeta function presented in Eq. \eqref{effP} is defined as 
 \begin{equation}
\zeta(s)=\sum_{N}\lambda_{N}^{-s},
\label{48}
\end{equation}
where $\lambda_{N}$ is the spectrum of eigenvalues associated to the self-adjoint elliptic operator $\Delta=-\square+\frac{R}{6}+V^{\prime\prime}_{cl}(\Phi)$ and $N$ stands for the set of quantum numbers associated to the quantum field eigenfunctions $\varphi$ of the operator $\Delta$.  Note that we used the shorthand notation 
$V^{\prime\prime}_{cl}(\Phi)=\dfrac{d^{2} V_{cl}(\Phi)}{d\Phi^2}$. Note also that the zeta function \eqref{48} relies on the complex parameter $s$ which is defined for Re$(s)>1$. Evidently, an analytic continuation to the whole complex $s$ plane can be obtained for the zeta function, including, in particular, $s=0$. Therefore, the regularized one-loop correction for the effective potential can be obtained by using the zeta function as defined in \eqref{48}.

\begin{figure}[!htb]
	\centering
					\includegraphics[width=0.6\textwidth]{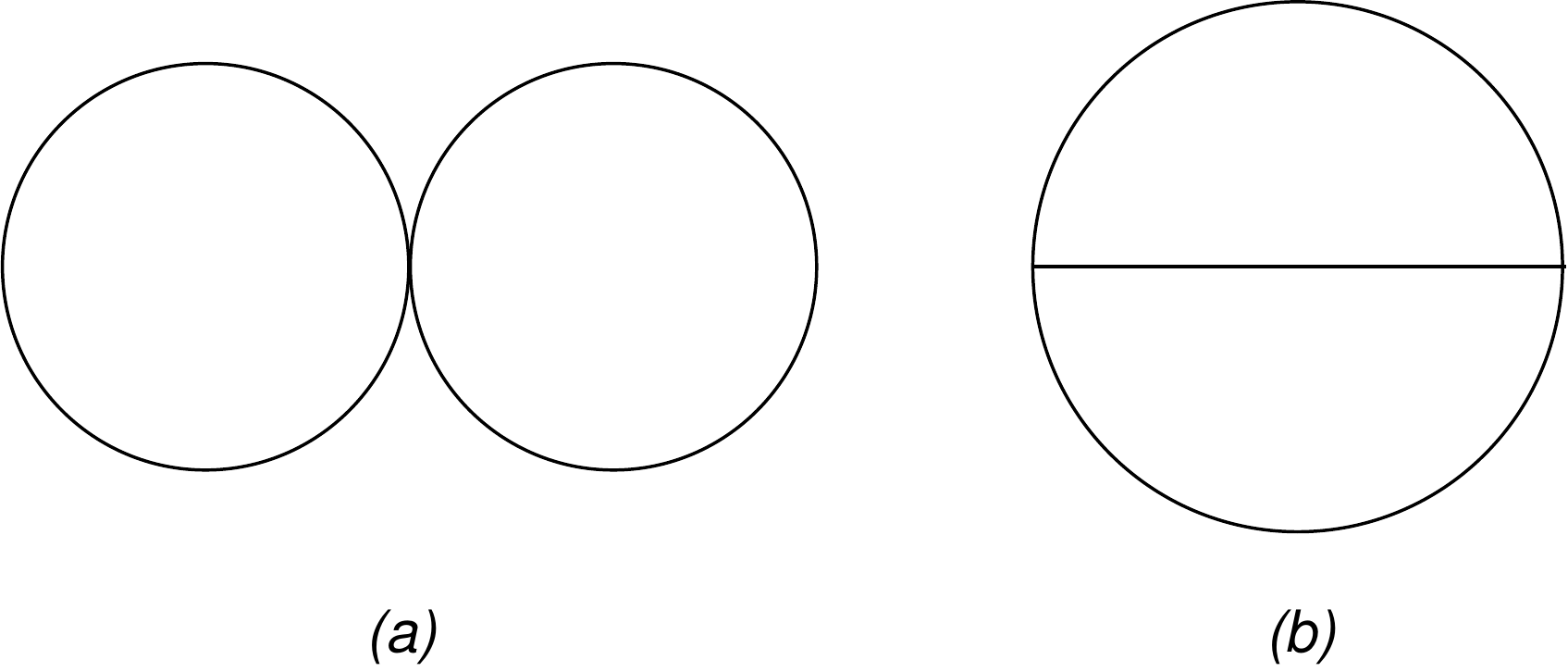}
	\caption{The figure displays the two graphs contributing to the effective potential}
	\label{fig2}
\end{figure}

To evaluate the two-loop contribution to the effective potential we will proceed in a different way than the one used for the one-loop contribution. The reason is  that only two graphs contribute to the effective action. So, it is more convenient to use the diagrammatic method rather than evaluate explicitly the full two-loop contribution to the effective potential. In this sense, Fig. \ref{fig2} displays the graphs which contribute to the two-loop effective potential. They yield the following result: 
\begin{equation}
V^{(2)}(\Phi)=\frac{\lambda}{8}S_{1}(\Phi)-\frac{\lambda^2}{12}\Phi^2 S_{2}(\Phi),
\label{two}
\end{equation}
where $S_{1}(\Phi)$ and $S_{2}(\Phi)$ are the contributions from the diagrams \ref{fig2}(a) and \ref{fig2}(b), respectively. In order to know the exact expressions for $S_{1}(\Phi)$ and $S_{2}(\Phi)$ we need to specify the topology of the spacetime. In our case, we will consider spacetimes with non-trivial topology and, as a consequence, the Feynman rules are similar to the ones for field theory at finite temperature, as remarked in Ref. \cite{toms2}. Furthermore, as we will deal with the vacuum state, $\Phi =0$, the second term of the r.h.s of \eqref{two} does not contribute to our results. An explicit form for $S_{1}(\Phi)$ will be given in the next (sub)-sections where we shall examine two different spacetimes with non-trivial topology. The first of them is the quasi-periodically identified Minkowski spacetime and the second one is the half-Einstein Universe.

Renormalization conditions necessary to obtain a renormalized effective potential should yet be considered \cite{toms2}. However, as we have only a renormalization constant, this implies that only one renormalization condition should be held. It can be taken in analogy to Coleman-Weinberg and fix the coupling-constant at some mass scale $M$, i.e.,

\begin{equation}
\frac{d^{4}V_{\text{eff}}}{d\Phi^4}\bigg|_{\Phi=M}=\lambda(M).
\label{NC}
\end{equation}   
Such a condition enables us to eliminate the dependence on $\mu$ in the one-loop contribution to the effective potential \eqref{effP}.

Now, in order to examine the generation of the topological mass it is necessary to impose the following equation:
\begin{equation}
\frac{d^{2}V_{\text{eff}}}{d\Phi^2}\bigg|_{\Phi=v}=m^2,
\label{TMass}
\end{equation} 
where $m$ represents the topological mass and $v$ is the value of $\Phi$ minimizing the effective potential and hence satisfying the extremum condition
\begin{equation}
\frac{dV_{\text{eff}}}{d\Phi}\bigg|_{\Phi=v}=0.
\end{equation}
Thus if the sign of the r.h.s. of Eq. (\ref{TMass}) is positive we conclude that $v$ is a minimum of the potential, otherwise is a maximum. 

\subsection{Flat spacetime: massless scalar field under quasi-periodic condition}
The procedure to calculate the values of energy of a massless scalar field under the quasi-periodic condition in the four-dimensional Minkowski spacetime is quite simple \cite{Feng:2013zza}. We can then apply this fact to calculate the eigenvalues $v_n$ corresponding to a massless scalar field eigenfunction $\varphi_n$ associated to the operator  $\Delta=-\square+\frac{R}{6}+V^{\prime\prime}_{cl}(\Phi)$, as described in the previous section. One should remind, though, that the $\Box$-operator is defined in the four-dimensional Euclidean space and time. Thereby, by subjecting a massless scalar field $\varphi$ to the quasi-periodic condition \cite{Feng:2013zza}
\begin{eqnarray}
\varphi(t,x,y,z+L) = e^{2i\pi\beta}\varphi(t,x,y,z),
\label{QPBC}
\end{eqnarray}
one finds the following eigenvalues:
\begin{eqnarray}
v_n = k^2+ \frac{4\pi^2}{L^2}\left(n+\beta\right)^2 + \frac{\lambda}{2}\Phi^2,
\end{eqnarray}
where $R=0$ in flat spacetime, $ k^2 = k_t^2 + k_x^2 + k_y^2 $, $n=0,\pm 1, \pm 2,...$, and $0\le \beta\le 1$. That is, the $z$-coordinate has been compactified into a length $L$ in order to impose on the field the quasi-periodic condition \eqref{QPBC} on the field. Moreover, one should notice that the case $\beta=0$ represents an untwisted scalar field and the case $\beta=\frac{1}{2}$ a twisted one. Both of these cases have already been previously considered in Ref. \cite{toms2}. Here, we want to consider a more general situation expressed by means of the parameter $\beta$.

The generalized zeta function is, thus, written as
\begin{eqnarray}
\zeta(s) = \frac{V_3}{(2\pi)^3}\sum_{n=-\infty}^{+\infty}\int d^{3}k\left[k^2 + \frac{4\pi^2}{L^2}\left(n+\beta\right)^2 + \frac{\lambda}{2}\Phi^2\right]^{-s}, 
\end{eqnarray}
where $V_3$ is the continuum volume associated to the dimensions $t,x,y$ and $d^3k=dk_tdk_xdk_y$. The integral above can be evaluated by using the relation 
\begin{equation}
\frac{1}{\omega^{s}}=\frac{2}{\Gamma(s)}\int_{0}^{\infty}d\tau\tau^{2s-1}e^{-\omega\tau^2}.
\label{rel}
\end{equation}
By doing so we find
\begin{eqnarray}
\zeta(s) = \frac{V_3}{(2\pi)^3}\frac{\pi^{\frac{3}{2}}\Gamma\left(s-\frac{3}{2}\right)}{\Gamma(s)}w^{3-2s}\sum_{n=-\infty}^{+\infty}\left[\left(n+\beta\right)^2 + \nu^2\right]^{\frac{3}{2}-s},
\label{GZF1}
\end{eqnarray}
where $\nu^2 = \frac{\lambda}{2w^2}\Phi^2$ and $w = \frac{2\pi}{L}$. The sum in $n$ considered in Eq. \eqref{GZF1} has been worked out before in Ref. \cite{Feng:2013zza} so that we will simply quote the result here. It is given by
\begin{eqnarray}
\sum_{n=-\infty}^{+\infty}\left[\left(n+\beta\right)^2 + \nu^2\right]^{-s} &=& \pi^{\frac{1}{2}}\nu^{1-2s}\frac{\Gamma\left(s-\frac{1}{2}\right)}{\Gamma(s)} + \frac{4\pi^s\nu^{\frac{1}{2}-s}}{\Gamma(s)}\nonumber\\
&\times&\sum_{k=1}^{\infty}k^{s-\frac{1}{2}}\cos(2\pi k\beta)K_{\left(\frac{1}{2}-s\right)}(2\pi k\nu),
\label{sum1}
\end{eqnarray}
where $K_{\gamma}(x)$ is the Macdonald function. Thereby, the generalized zeta function in Eq. \eqref{GZF1} becomes
\begin{eqnarray}
\zeta(s) &=& \frac{V_3}{(2\pi)^3}\pi^{\frac{3}{2}+\bar{s}}w^{-2\bar{s}}\nu^{1-2\bar{s}}\left[\pi^{\frac{1}{2}-\bar{s}}\frac{\Gamma\left(\bar{s}-\frac{1}{2}\right)}{\Gamma\left(\bar{s}+\frac{3}{2}\right)} + \frac{4(2\pi)^{\frac{1}{2}-\bar{s}}}{\Gamma\left(\bar{s}+\frac{3}{2}\right)}\sum_{k=1}^{\infty}\cos(2\pi k\beta)f_{\left(\frac{1}{2}-\bar{s}\right)}(2\pi k\nu)\right],\nonumber\\
\label{zeta1}
\end{eqnarray}
where $\bar{s}=s-\frac{3}{2}$ and the function $f_\gamma(x)$ is defined as
\begin{eqnarray}
f_\gamma(x) = \frac{K_{\gamma}(x)}{x^{\gamma}}.
\label{besselk1}
\end{eqnarray}
Note that, in what follows, the limit $s\rightarrow 0$  is equivalent to the limit $\bar{s}\rightarrow -\frac{3}{2}$. Thus, in this case, the one-loop correction to the effective potential given by Eq. \eqref{effP} is written in terms of
\begin{eqnarray}
\zeta(0) = \frac{V_3L}{2(2\pi)^4}\pi^2b^4,
\label{zeta0}
\end{eqnarray}
and
\begin{eqnarray}
\zeta'(0) = \frac{V_3L}{(2\pi)^4}\left[\frac{\pi^2}{4}b^4(3-4\ln(b)) + 4w^2b^2\sum_{k=1}^{\infty}\frac{1}{k^2}K_{2}\left(\frac{2\pi kb}{w}\right)\cos(2 k\pi\beta)\right],
\label{zetad0}
\end{eqnarray}
where $b^2=\frac{\lambda}{2}\Phi^2$ and $V_3L$ is the four-dimensional volume of the Euclidean space and time. Then, by using Eqs. \eqref{zeta0} and \eqref{zetad0}, the one-loop correction can be written as
\begin{eqnarray}
V_{\text{eff}}(\Phi) &=& \frac{\lambda}{4!}\Phi^4 + \frac{C}{4!}\Phi^4 - \frac{1}{32\pi^4}\left[\frac{\pi^2b^4}{2}\ln(\mu^2) + \frac{\pi^2}{4}b^4(3-4\ln(b))\right.\nonumber\\
&&\left.+ 4w^2b^2\sum_{k=1}^{\infty}\frac{1}{k^2}K_{2}\left(\frac{2\pi kb}{w}\right)\cos(2k\pi\beta)\right].
\label{effP1}
\end{eqnarray}
The normalization constant $C$ is obtained, after taking $L\rightarrow\infty$, by using Eq. \eqref{NC}. Then, it provides
\begin{eqnarray}
\frac{C}{4!} =\frac{\lambda^2}{256\pi^2}\ln(\mu^2) - \frac{\lambda^2}{96\pi^2} - \frac{\lambda^2}{256\pi^2}\ln\left(\frac{\lambda M^2}{2}\right),
\label{rc}
\end{eqnarray}
which does not depend on $L$ and, therefore, is the infinite $L$ contribution (in the finite temperature case based on similar summations, the analogous contribution is the zero-temperature one). The use of the normalization constant into Eq. \eqref{effP1} allows us to write the renormalized effective potential with one-loop correction as
\begin{eqnarray}
V^{\text{R}}_{\text{eff}}(\Phi) = \frac{\lambda}{4!}\Phi^4 +
\frac{\lambda^2\Phi^4}{(16\pi)^2}\left[\ln\left(\frac{\Phi^2}{M^2}\right)-\frac{25}{6}\right] - \frac{b^2}{2\pi^2L^2}\sum_{k=1}^{\infty}\frac{1}{k^2}K_{2}\left(kbL\right)\cos(2k\pi\beta).
\label{effP1R}
\end{eqnarray}
It is clear that, by taking the limit $L\rightarrow\infty$, we obtain the well known Coleman-Weinberg effective potential, confirming the consistency of our results.

The state $\Phi=0$, characterizing the ground state at the tree-level of the effective potential, provides the Casimir energy density through the expression \eqref{effP1R}, that is,
\begin{eqnarray}
V^{\text{R}}_{\text{eff}}(\Phi = 0) = - \frac{1}{\pi^2L^4}\sum_{k=1}^{\infty}\frac{1}{k^4}\cos(2k\pi\beta).
\label{effP1Rzero}
\end{eqnarray}
This result is definitely finite and this can be better seen by using the Bernoulli polynomials 
\begin{eqnarray}
B_{2k}(x) = \frac{(-1)^{k-1}2(2k)!}{(2\pi)^{2k}}\sum_{n=1}^{\infty}\frac{\cos(2\pi n x)}{n^{2k}},\qquad\qquad 0\leq x \leq 1,
\label{Bern}
\end{eqnarray}
where $k=0,1,2,...$\;. Thereby, the expression \eqref{effP1Rzero} becomes
\begin{eqnarray}
V^{\text{R}}_{\text{eff}}(\Phi = 0)= \frac{\pi^2}{3L^4}B_4(\beta)=\frac{\pi^2}{3L^4}\left(\beta^4 - 2\beta^3 + \beta^2 - \frac{1}{30}\right),
\label{effP1Rzero2}
\end{eqnarray}
which is the same result obtained in Ref. \cite{Feng:2013zza}, as it should be. It is also worth mentioning the fact that the expression \eqref{effP1Rzero2}, provides the correct results for the untwisted $(\beta=0)$ and twisted $\left(\beta=\frac{1}{2}\right)$ cases. Both of these cases were separately analyzed in Ref. \cite{toms2}.

\begin{figure}[!htb]
\begin{center}
\includegraphics[width=0.45\textwidth]{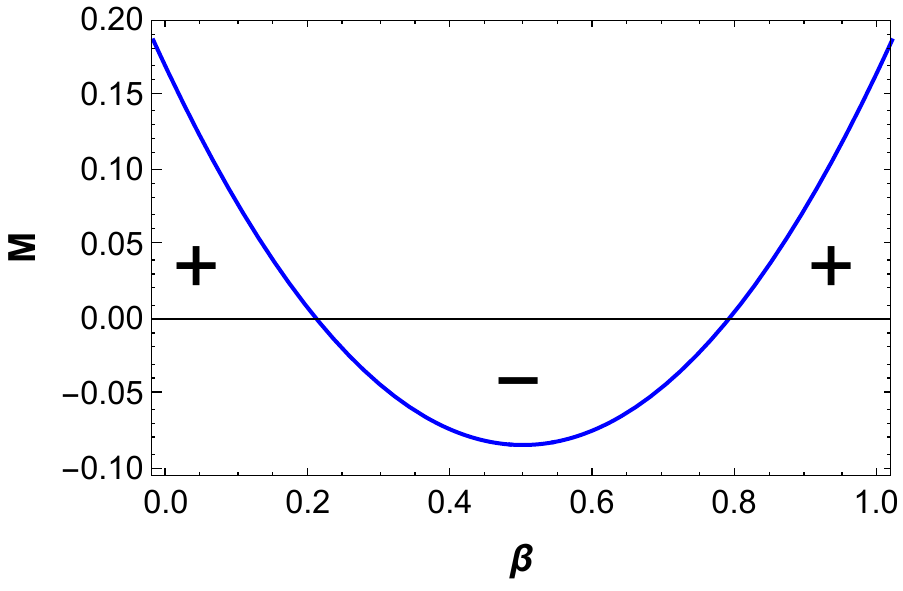}
\caption{\small{Plot of the dimensionless quantity $M=\frac{4L^2m^2}{\lambda}$ in terms of the parameter $\beta$, according to Eq. \eqref{topmass2}.}}
\label{f1}
\end{center}
\end{figure}

The stability of the ground state is examined by means of the condition \eqref{TMass}. If it is negative, $\Phi=0$ is a local maximum, otherwise is a local minimum and provides a topological positive mass. Thus, by using \eqref{TMass} and making $\Phi=0$  we get
\begin{eqnarray}
m^2 =\frac{\lambda}{4L^2}B_2(\beta)=\frac{\lambda}{4L^2}\left(\beta^2 - \beta + \frac{1}{6}\right).
\label{topmass2}
\end{eqnarray}
To the best of our knowledge, this expression has been obtained for the first time here. It is evident that the result in Eq.  \eqref{topmass2} is not positive for all values of $\beta$ between zero and one. In fact, it is consistent with the results for the untwisted $(\beta=0)$ and twisted $\left(\beta=\frac{1}{2}\right)$ cases obtained in Ref. \cite{toms2}. For the twisted case, the expression above is negative and may indicate instability, as pointed out in Ref. \cite{toms3}. In Fig.\ref{f1} we have plotted the dimensionless quantity $M=\frac{4L^2m^2}{\lambda}$ as a function of the parameter $\beta$. We can see that it is positive only for $\beta<0.2$ and for $\beta>0.8$, making the tree-level vacuum state $\Phi=0$ stable. All the other values of $\beta$, including the value that provides the twisted case, give a negative result for the expression \eqref{topmass2}.

%

Let us now calculate the two-loop contribution to the vacuum energy density. In order to do that, we have to make use of the expression in Eq. \eqref{two}, taking $\Phi=0$. In this case, there is only contribution from the first term on the r.h.s. of  \eqref{two}. Thus, from the graphs in Fig.\ref{fig2} we have
\begin{eqnarray}
S_1(\Phi) &=& \left\{\sum_{n=-\infty}^{+\infty}\frac{1}{L}\int\frac{d^{3}k}{(2\pi)^3}\left[k^2 + \frac{4\pi^2}{L^2}\left(n+\beta\right)^2 + \frac{\lambda}{2}\Phi^2\right]^{-s}\right\}^2\nonumber\\
&=& \left(\frac{1}{V_3L}\zeta(s)\right)^2,
\label{intcont}
\end{eqnarray}
where we have to take $s=1$. Note that in Ref. \cite{toms2} dimensional regularization was used to calculate this integral. However, we would like here to use the  zeta function regularization, that is why we have introduced the parameter $s$ in Eq. \eqref{intcont}. Moreover, if we naively take $s=1$ in the expression in Eq. \eqref{zeta1} to calculate the integral  \eqref{intcont}, we will find a divergence coming from the first term on the r.h.s. Nevertheless, a thorough inspection of the zeta function in Eq. \eqref{zeta1} will show us that, when divided by $V_3L$, its first term does not depend on the length $L$ and, therefore, can be dropped, as usually it is done in similar calculations. The result of the integral \eqref{intcont} in the ground state, thus, is found to be
\begin{eqnarray}
S_1(\Phi=0)=\left(\frac{1}{V_3L}\zeta_{\text{R}}(1)\right)^2 = \frac{1}{4L^4}B_2^2(\beta),
\label{intcont2}
\end{eqnarray}
where $\zeta_{\text{R}}(1)$ means that the expression \eqref{zeta1} was taken at $s=1$, dropping its first term. Finally, as it follows from Eq. \eqref{two}, the two-loop contribution to the effective potential is given by
\begin{eqnarray}
V^{(2)}(\Phi=0)=\frac{\lambda}{8}S_1(\phi=0)=\frac{1}{32L^4}\left(\beta^2 - \beta + \frac{1}{6}\right)^2.
\label{twoloop}
\end{eqnarray}
Note that, in the general expression above, when one considers the untwisted and twisted cases the results coincide with the ones found in Ref. \cite{toms2}. This shows the consistency of our regularization method by making use of the zeta function \eqref{zeta1}.

\subsection{Massless scalar field in the `half-Einstein Universe'}
%
The static Einstein Universe has already been considered within the context of the Casimir effect previously in several works \cite{toms2, Ford:1978zt, FORD197989, PhysRevD.83.104042, Bezerra:2011nc}. In Ref.  \cite{toms2}, for instance, the author considered loop corrections to the ground state energy density of a massless scalar field, he also studied generation of topological mass. Here, on the other hand, we wish to consider the static Einstein Universe subjected to a Dirichlet boundary condition. This has been called by the authors in Refs.  \cite{Bayin:1993yq, Ozcan:2001cr, Kennedy_1980} as `half-Einstein Universe'. The eigenvalues of energy of a massless scalar field in this scenario have also been found in Ref. \cite{Bayin:1993yq} and, in our case, we can make use of this. Thus, the eigenvalues  of a massless scalar field, subjected to Dirichlet boundary condition, and which is also eigenvalues of the operator $\Delta=-\square+\frac{R}{6}+V^{\prime\prime}_{cl}(\Phi)$, are given by 
\begin{eqnarray}
v_n = \frac{\lambda}{2}\Phi^2 +k_t^2+\frac{n^2}{a^2},
\label{Dir}
\end{eqnarray}
where the curvature scalar $R=\frac{6}{a^2}$ has been used, $a$ is the constant scale factor and $k_t$ is the momentum associated to the $t$-coordinate. Note that the $\Box$-operator is defined for the geometry of the static Einstein Universe.

The generalized zeta function is, then, built out from \eqref{Dir} and \eqref{48}. It is written as
\begin{equation}
\zeta(s) = \left(\frac{L_t}{2\pi}\right)\int_{-\infty}^{\infty}dk_t\sum_{n=1}^{\infty}d_D(n)\left[ \frac{\lambda}{2}\Phi^2+\frac{n^2}{a^2}+k_t^2\right]^{-s},
\label{zNeu}
\end{equation}
where $L_t$ is a parameter with dimension of length and $d_D(k)$ is the degeneracy given by $d_D(n) = \frac{1}{2}n(n-1)$, with $n=1,2,3,...$\; \cite{Bayin:1993yq}. In the case when Neumann boundary condition is used, the eigenvalues are the same as \eqref{Dir} but with a degeneracy given by $d_N(n) = \frac{1}{2}n(n+1)$. Thereby, using again the relation in Eq. \eqref{rel} we get
\begin{equation}
\zeta(s) =\sqrt{\pi} \left(\frac{L_t}{2\pi}\right)\frac{\Gamma(s-\frac{1}{2})}{\Gamma(s)}a^{2s-1}\sum_{n=1}^{\infty}\frac{1}{2}n(n-1)\left[ \nu^2+n^2\right]^{\frac{1}{2}-s},
\label{zNeu2}
\end{equation}
where $\nu^2=a^2\frac{\lambda}{2}\Phi^2$. 

In order to perform the sum over $n$ in Eq. \eqref{zNeu2} it is convenient to break it into two contributions: one with the multiplicative factor given by $n^2$ and the other with the multiplicative factor given by ($-n$). Let us start with the first one. This contribution is written as
\begin{eqnarray}
\zeta_I(s) &=&\sqrt{\pi} \left(\frac{L_t}{2\pi}\right)\frac{\Gamma\left(s-\frac{1}{2}\right)}{2\Gamma(s)}a^{2s-1}\sum_{n=1}^{\infty}n^2\left[ \nu^2+n^2\right]^{\frac{1}{2}-s}\nonumber\\
&=&\sqrt{\pi} \left(\frac{L_t}{2\pi}\right)\frac{\Gamma(\bar{s})}{2\Gamma\left(\bar{s}+\frac{1}{2}\right)}a^{2\bar{s}}\left[\zeta_{EH}(\bar{s}-1,\nu)-\nu^2\zeta_{EH}(\bar{s},\nu)\right],
\label{zetaI}
\end{eqnarray}
where $\zeta_{EH}(\bar{s},\nu)$ is the Epstein-Hurwitz zeta function \cite{Elizalde:1994gf, Bezerra:2017zqq} and $\bar{s}=s-\frac{1}{2}$. Note that, in what follows, taking the limit $s\rightarrow 0$ is equivalent to the limit $\bar{s}\rightarrow -\frac{1}{2}$.

The Epstein-Hurwitz  zeta function is defined as \cite{Elizalde:1994gf, Bezerra:2017zqq} 
\begin{eqnarray}
\zeta_{EH}(s,\nu) = \sum_{n=1}^{\infty}(n^2 + \nu^2)^{-s},
\label{dezeta}
\end{eqnarray}
where Re($s)>1/2$ and $\nu^2\ge 0$. A very useful expression for this sum providing an analytical continuation for other values of $s$ is given by \cite{Elizalde:1994gf, Bezerra:2017zqq} 
\begin{eqnarray}
\zeta_{EH}(s,\nu) = -\frac{\nu^{-2s}}{2} + \frac{\sqrt{\pi}}{2}\frac{\Gamma\left(s-\frac{1}{2}\right)}{\Gamma(s)}\nu^{1-2s} + \frac{2^{1-s}(2\pi)^{2s-\frac{1}{2}}}{\Gamma(s)}\sum_{k=1}^{\infty}k^{2s-1}f_{\left(s-\frac{1}{2}\right)}(2\pi k\nu),
\label{ACzeta}
\end{eqnarray}
where $f_{\gamma}(x)$ has been defined in Eq. \eqref{besselk1}. By using \eqref{ACzeta} in Eq. \eqref{zetaI} we are able to obtain 
\begin{eqnarray}
\zeta_I(0) = \pi\left(\frac{L_t}{2\pi}\right)\frac{\nu^4}{16a},
\label{zetazero}
\end{eqnarray}
and
\begin{eqnarray}
\zeta'_I(0) = \pi\left(\frac{L_t}{2\pi}\right)\frac{\nu^4}{32a}\left\{3+4\ln\left(\frac{a}{\nu}\right)-64\sum_{n=1}^{\infty}\left[f_1(2\pi n\nu) + 3f_2(2\pi n\nu)\right]\right\}.
\label{zetadezero}
\end{eqnarray}

Let us now turn our attention to the second contribution. From Eq. \eqref{zNeu2} we can see that it is written as
\begin{eqnarray}
\zeta_{II}(s) =-\sqrt{\pi} \left(\frac{L_t}{2\pi}\right)\frac{\Gamma\left(s-\frac{1}{2}\right)}{2\Gamma(s)}a^{2s-1}\sum_{n=1}^{\infty}n\left[ \nu^2+n^2\right]^{\frac{1}{2}-s}.
\label{zetaII}
\end{eqnarray}
In order to perform the sum in $n$ exhibited in the above expression we can apply the Abel-Plana formula given by \cite{PhysRevD.83.104042, Bezerra:2011nc}
\begin{eqnarray}
\sum_{n=1}^{\infty}F(n) = -\frac{1}{2}F(0) + \int_{0}^{\infty}F(t)dt + i\int_{0}^{\infty}dt\frac{[F(it)-F(-it)]}{e^{2\pi t}-1},
\label{ABF}
\end{eqnarray}
where, in our case, $F(n) = n\left[ \nu^2+n^2\right]^{\frac{1}{2}-s}$. Taking this into account, we see that the first term on the r.h.s of \eqref{ABF} is zero. Furthermore, the second and third terms are, respectively, given by
\begin{eqnarray}
 \int_{0}^{\infty}t\left[ \nu^2+t^2\right]^{-\bar{s}}dt = \frac{\nu^{2-2\bar{s}}}{2(\bar{s}-1)},
 \label{int1}
\end{eqnarray}
and
\begin{eqnarray}
i\int_{0}^{\infty}dt\frac{[F(it)-F(-it)]}{e^{2\pi t}-1} &=& -A(\bar{s})\sum_{k=1}^{\infty}\int_{\nu}^{\infty}dt t(t^2-\nu^2)^{-2\bar{s}}e^{-2\pi kt}\nonumber\\
&=& -A(\bar{s})\sqrt{\pi}\sum_{k=1}^{\infty}kf_{\left(\frac{3}{2}-\bar{s}\right)}(2\pi k\nu),
 \label{int2}
\end{eqnarray}
where $A(s)=2^{\frac{3}{2}-\bar{s}}\nu^{3-2\bar{s}}i^{-2\bar{s}}\left[1+(-1)^{-2\bar{s}}\right]$ and we have used the identities $\sum\limits_{k=1}^{\infty}e^{-2\pi kt}=e^{2\pi t}-1$ and 
\begin{eqnarray}
\left[\nu^2 + (\pm i t)^2\right]^{-\bar{s}}=\left\{ \begin{array}{l}[\nu^2 - t^2]^{-\bar{s}},\,\,\;\;\;\;\;\;\;\;\;\;\,\mathrm{for}\,\, \nu>t\,,\\
(\pm i)^{\bar{s}}[t^2 - \nu^2]^{-\bar{s}},\,\,\,\,\,\,\mathrm{for}\,\, \nu<t\,.
\end{array}\right.
\label{iden}
\end{eqnarray}
Collecting the results in Eqs. \eqref{int1} and \eqref{int2}, one writes the expression in Eq. \eqref{zetaII} as
\begin{eqnarray}
\zeta_{II}(s) =-\sqrt{\pi} \left(\frac{L_t}{2\pi}\right)\frac{\Gamma(\bar{s})}{2\Gamma\left(\bar{s}+\frac{1}{2}\right)}a^{2\bar{s}}\left[\frac{\nu^{2-2\bar{s}}}{2(\bar{s}-1)} - A(\bar{s})\sqrt{\pi}\sum_{k=1}^{\infty}kf_{\left(\frac{3}{2}-\bar{s}\right)}(2\pi k\nu)\right].
\label{zetaII1}
\end{eqnarray}
This expression allows to take the limit $s\rightarrow 0$, or equivalently $\bar{s}\rightarrow -\frac{1}{2}$. This gives 
\begin{eqnarray}
\zeta_{II}(0) = -\pi\left(\frac{L_t}{2\pi}\right)\frac{\nu^4}{3a},
\label{zetazeroII}
\end{eqnarray}
and $\zeta'_{II}(0) =0$. We can see that the latter does not give any contribution to the one-loop effective potential. The only contribution would come from Eq. \eqref{zetazeroII}, but this term enters as a factor accompanying $\ln(\mu^2)$ and will be absorbed by the renormalization constant $C$, as we shall see below. Note that adopting the Neumann boundary condition would only change the sign of \eqref{zetazeroII}. 

The effective potential with one-loop correction now can be obtained using Eq. \eqref{effP}, and the results in Eqs. \eqref{zetazero}, \eqref{zetadezero} and \eqref{zetazeroII}, that is,
\begin{eqnarray}
V_{\text{eff}}(\Phi) &=& \frac{\Phi^2}{2a^2} + \frac{\lambda}{4!}\Phi^4 + \frac{C}{4!}\Phi^4\nonumber\\
&-& \frac{1}{128\pi^2}\left\{-\frac{13b^4}{3}\ln(\mu^2) + \frac{3b^4}{2} - 2b^4\ln\left(b\right) -32b^4 \sum_{k=1}^{\infty}\left[f_1(2\pi n\nu) + 3f_2(2\pi n\nu)\right]\right\},
\label{effPE1}
\end{eqnarray}
where we have used $R=\frac{6}{a^2}$, $b^2=\frac{\lambda}{2}\Phi^2$ and the four-dimensional spacetime volume of the Einstein Universe $V = 2\pi^2L_t a^3$ \cite{toms2}. Had we used Neumann boundary condition, the first term in brackets on the r.h.s of \eqref{effPE1} would be $\frac{19b^4}{3}\ln(\mu^2)$ as a consequence of the change in sign of \eqref{zetazeroII}. Ultimately, this will not contribute to the renormalized one-loop effective potential, providing that the ground state energy at one-loop is not affected by whether the boundary condition is Dirichlet or Neumann.

The normalization condition expression in Eq. \eqref{NC} can by applied for the effective potential \eqref{effPE1}, after taking $a\rightarrow\infty$. This provides 
\begin{eqnarray}
\frac{C}{4!} =-\frac{13\lambda^2}{1536\pi^2}\ln(\mu^2) + \frac{3\lambda^2}{1024\pi^2} - \frac{\lambda^2}{512\pi^2}\left[\ln\left(\frac{\lambda M^2}{2}\right)+ \frac{25}{6}\right].
\label{rcE}
\end{eqnarray}
Furthermore, by substituting this into Eq. \eqref{effPE1} we obtain the one-loop  renormalized effective potential in the form
\begin{eqnarray}
V^{\text{R}}_{\text{eff}}(\Phi) = \frac{\Phi^2}{2a^2} + \frac{\lambda}{4!}\Phi^4 - \frac{25}{3072\pi^2}\lambda^2\Phi^4 + \frac{\lambda^2\Phi^4}{512\pi^2}\ln\left(\frac{\Phi^2}{M^2}\right)  + \frac{\lambda^2\Phi^4}{16\pi^2} \sum_{k=1}^{\infty}\left[f_1(2\pi n\nu) + 3f_2(2\pi n\nu)\right].
\label{effPE1RR}
\end{eqnarray}
Thereby, the energy density of the ground state $\Phi=0$ follows from the renormalized one-loop effective potential. It is found to be
\begin{eqnarray}
V^{\text{R}}_{\text{eff}}(\Phi=0) &=&\frac{3}{32a^4\pi^6}\zeta(4)=
\frac{1}{960\pi^2a^4},
\label{effPE1R}
\end{eqnarray}
where $\zeta(4) = \frac{\pi^4}{90} $ is the known value of the Riemann zeta function  \cite{Elizalde:1994gf}. As we have mentioned before, the expression \eqref{effPE1R} is the same with no matter whether the boundary condition is Dirichlet or Neumann. Moreover, although we have obtained the result in Eq. \eqref{effPE1R} using loop corrections it had already been obtained previously by the authors in Ref.  \cite{Bayin:1993yq}. So, our result agrees with the one reported in \cite{Bayin:1993yq}.

What is new here, to the best of our knowledge, is the value of the topological mass obtained by the condition \eqref{TMass}. It gives 
\begin{eqnarray}
m^2 &=& \frac{1}{a^2} - \frac{\lambda}{32\pi^4a^2}\zeta(2)=
\frac{1}{a^2}\left(1-\frac{\lambda}{192\pi^2}\right),
\label{topmassE}
\end{eqnarray}
where $\zeta(2)=\frac{\pi^2}{6}$. It is interesting to mention that the numerical factor in the second term at the r.h.s of \eqref{topmassE} is half of the numerical factor of the one in Eq. (70) of Ref. \cite{toms2}. Since $\lambda$ is small, the expression \eqref{topmassE} is always positive.

The two-loop contribution to the vacuum energy can be obtained along the same lines as in the previous (sub)-section. The only contribution to the vacuum energy, obtained at $\Phi=0$, comes from the first term in the r.h.s of Eq.\eqref{two}. This term can be calculated with the help of the graphs in Fig.\ref{fig2}, i.e.,
\begin{eqnarray}
S_1(\Phi) &=& \left\{\sum_{n=1}^{+\infty}\frac{1}{2\pi^2a^3}\int\frac{dk_t}{(2\pi)}d_D(n)\left[\frac{n^2}{a^2} + k_t^2 + \frac{\lambda}{2}\Phi^2\right]^{-s}\right\}^2\nonumber\\
&=& \left(\frac{1}{2\pi^2a^3L_t}\zeta(s)\right)^2,
\label{intcont2a}
\end{eqnarray}
where we have to take $s=1$ and $\zeta(s)=\zeta_I(s)+\zeta_{II}(s)$ is given in terms of Eqs.  \eqref{zetaI}, \eqref{ACzeta} and \eqref{zetaII1}. Again,  if we naively take $s=1$ in the expression for the generalized zeta function, to calculate the integral  \eqref{intcont2a}, we will find a divergence coming from the second term on the r.h.s of \eqref{ACzeta}. However, once this term is divided by $2\pi^2a^3L_t$ we can see that it does not depend on $a$ and can be dropped. The result is then
\begin{eqnarray}
S_1(\Phi=0) &=& \left(\frac{1}{2\pi^2a^3\beta}\zeta_{\text{R}}(1)\right)^2=
\frac{1}{9216 a^4 \pi^4},
\label{twoloopa}
\end{eqnarray}
where $\zeta_{\text{R}}(1)$ means that the expression $\zeta(s)$ was taken at $s=1$, dropping the second term on the r.h.s of \eqref{ACzeta}. Thus, we finally have the two-loop contribution to the vacuum energy 
\begin{eqnarray}
V^{(2)}(\Phi=0) &=& \frac{\lambda}{8}S_1(\Phi=0)=
\frac{\lambda}{73728 a^4\pi^4}.
\label{twoloopE}
\end{eqnarray}
The numerical factor is one fourth of the numerical factor of Eq. (73) in Ref. \cite{toms2}. The two-loop contribution to the vacuum energy above also is the same both for Dirichlet and Neumann boundary conditions. To the best of our knowledge, the result in Eq. \eqref{twoloopE} has been for the first time obtained here.

   \section{Conclusions}
\label{secIII}

We have investigated the ground state energy (Casimir effect) in the $\lambda\phi^4$ theory non-minimally coupled to gravity. In addition, we have given a brief review of some aspects on path integral approach, emphasizing the quantum corrections to the effective potential up to the two-loop order and the mechanism of generation of topological mass by considering a quasi-periodically identified Minkowski spacetime and the half-Einstein universe, that is, an Einstein Universe where a massless scalar field propagates under Dirichlet boundary condition. We have found a renormalized effective potential in both cases considered, Eqs. \eqref{effP1R} and  \eqref{effPE1RR}. At the tree-level graph there is no ground state energy different from zero while considering one and two-loop corrections there are nonzero contributions.

In the quasi-periodically identified Minkowski spacetime case, a nonzero ground state energy at $\Phi=0$, Eq. \eqref{effP1Rzero2}, was obtained from the one-loop correction to the effective potential,  Eq. \eqref{effP1R}. This result obtained here by means of the effective potential has already been described before in the literature and shows the consistency of our approach. A nonzero contribution to the ground state energy at two-loop levels has also been obtained in Eq. \eqref{twoloopa} as well as the topological mass, Eq. \eqref{topmass2}, generated by the quasi-periodic condition \eqref{QPBC}. This new general result obtained here is consistent with previous results found earlier for the twisted and untwisted scalar fields. The behavior of the topological mass with respect to the phase $\beta$ is plotted in Fig.\ref{f1}.

On the other hand, in the half-Einstein Universe case, a nonzero ground state energy at $\Phi=0$, Eq. \eqref{effPE1R}, was also obtained from the one-loop correction to the effective potential,  Eq. \eqref{effPE1RR}. This result has already been obtained in literature by other method and, as in the case for the quasi-periodically identified Minkowski spacetime, shows the consistency of our approach. Moreover, the new results obtained in this case are the topological mass, Eq. \eqref{topmassE}, and the two-loop contribution to the ground state energy in Eq. \eqref{twoloopE}. A natural continuation of this work consist in carrying out the same analysis for other more involved backgrounds, for example, AdS space. This study is currently under way.


\section*{Acknowledgments}
We wish to thank F. S. Gama and A.Y. Petrov for important discussions and collaboration on related topics. P.J.P would like to thank the Brazilian agency CAPES financial support (PDE/CAPES grant, process 88881.171759/2018-01). H.F.S.M would like to thank CNPq for partial financial support (PQ/CNPq grants 305379/2017-8 and 430002/2018-1). G.Q.G would like to thank the Brazilian agencies CNPq and Fapesq-PB for the financial support (Fapesq-PB/CNPq grant, process 300354/2018-5).

\end{document}